\documentclass[prb,superscriptaddress,showpacs,twocolumn]{revtex4}

\usepackage{graphicx}
\usepackage{amsmath,amsthm,amssymb}

\begin{document}

\title{Dynamics of driven vortex-antivortex matter in superconducting films with a magnetic dipole array}

\date{\today}

\author{Cl\'essio L. S. Lima}
\affiliation{Departamento de F\'{\i}sica, Universidade Federal de Pernambuco, 50670-901 Recife-PE, Brazil}
\affiliation{Instituto de F\'{\i}sica, Universidade Federal do Rio de Janeiro, C.P. 68528, 21945-970, Rio de Janeiro-RJ, Brazil}

\author{Cl{\'e}cio C. de Souza Silva}
\affiliation{Departamento de F\'{\i}sica, Universidade Federal de Pernambuco, 50670-901 Recife-PE, Brazil}

\begin{abstract}
We investigate theoretically vortex-antivortex (v-av) matter moving in thin superconducting films with a regular array of in-plane magnetic dipoles. Our model considers v-av pair creation induced by the local current density generated by the magnetic texture and the transport current and simulates the dynamics of vortices and antivortices by numerical integration of the Langevin equation of motion. Calculations of the transport properties at zero applied field show a strong dependence of the v-av dynamics on the current intensity and direction. The dynamics of the v-av matter is characterized by a series of creation and annihilation processes, which reflect on the time dependence of the electrical field, and by guided motion, resulting in a zero-field transverse resistance. 
\end{abstract}


\maketitle

A small magnetic dot placed in the vicinity of a thin superconducting film has the ability of creating
vortex-antivortex (v-av) pairs in the superconductor and keep them from annihilating each other.\cite{Erdin02,Milosevic02} Thanks to this, phases of vortex matter in which vortices and antivortices coexist can be stabilized in bilayers made of thin superconducting films and arrays of nanomagnets, even in the absence of macroscopic magnetic fields. This topic has attracted a great deal of attention lately. Several calculations using the Ginzbug-Landau approach have been carried out for arrays of polarized magnetic dots on top of a superconducting film, predicting a great variety of equilibrium phases of vortices and anti-vortices.\cite{Priour03,Milosevic0405}  

Most early experimental work on superconductor-nanomagnet hybrids were aimed at the flux pinning properties of the magnetic dots.\cite{Schuller,VanBael,Villegas08} Recently, experiments on superconducting films with polarized magnetic dot arrays corroborate the existence of vortex-antivortex phases in these structures and show that such phases induce novel phenomena such as the field-induced superconductivity and the zero-magnetic-field ratchet effect.\cite{Lange03,Souza07} 

A problem that has received much less attention is the existence in these systems of dynamical states of vortex matter in which vortices and antivortices coexist. These states may result when a transport current is applied to the film, and the v-av matter moves. In this paper we study this problem theoretically. We expect that during the motion v-av pairs annihilate and that new pairs are created. The fundamental question is how these processes, together with vortex-vortex and vortex-nanomagnet interactions, determine the properties of the moving vortex matter. Here we address this point by using numerical simulations of a simple model for vortices interacting with a periodic array of permanent point dipoles.

\begin{figure}[b]
\centerline{\includegraphics[scale=0.85]{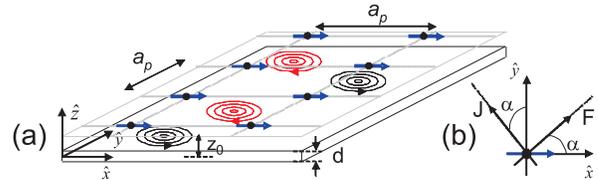}}
\caption{(Color online) (a) Schematic view of the in-plane dipole array. (b) Definition of the transport current {\bf J} and driving force {\bf F} orientations. }
\label{fig.system}
\end{figure}

The details of our model is as follows. The bilayer comprises a thin superconducting film of thickness $d$, penetration depth $\lambda\gg d$, and coherence length $\xi<d$. A square array of magnetic dipoles, with magnetic moment ${\bf m}$ and unit cell of dimensions $a_p\times a_p$ (Fig.~\ref{fig.system}),
is placed above the film, at a distance $z_0>d/2$ from it. Experimentally, the space between the superconducting film and the dipole layer is usually filled with an insulating buffer layer to prevent proximity effects. Here we assume that the dipoles are polarized in plane, parallel to the $x$ direction (${\bf m}=m\hat{\bf x}$). The total energy of the vortex matter is given by
\begin{equation}
 E=\sum_{i,j}\,q_iq_j\,\Big[\frac{1}{2}U_{\rm vv}({\bf r}_{ij})+
 E_c\delta_{i,j}\Big]+\sum_j q_j U_{\rm vm}({\bf r}_j),
\label{eq.et1}
\end{equation}
where $q$ is  the vorticity ($q=1$ for vortices, $q=-1$ for antivortices). Here we assume that no external magnetic field is applied. In other words, the  vortex-matter ``neutrality'' condition, $\sum_j q_j=0$, must be satisfied. 
$U_{vm}({\bf r})$ is the usual interaction energy between a vortex located at ${\bf r}$ and the dipole array calculated in the London limit.\cite{Carneiro05} 
The interaction energy $U_{\rm vv}({\bf r}_{ij})$ between an arbitrary pair of vortices at positions ${\bf r}_i$ and ${\bf r}_j$ is calculated following Clem's variational approach for the order parameter (${\bf r}_{ij}={\bf r}_i-{\bf r}_i$).\cite{Clem74} In the limit $\lambda\gg d$ considered here, this gives $U_{\rm vv}({\bf r})= \epsilon_0d\ln(|{\bf r}_{ij}|^2 + 2\xi_c^2)$, where $\epsilon_0 = \phi_0^2/4\pi\mu_0\lambda^2$ and $\xi_c$ is a variational parameter (in general, $\xi_c \sim \xi(T)$). The vortex core energy can be calculated straightforwardly by using Clem's trial order parameter to compute the gain in condensation energy due to an isolated vortex, which results in $E_c=0.375\epsilon_0$. $-E_c$ acts as a chemical potential of the vortex system and thus plays an important role in the nucleation of vortex-antivortex pairs.\cite{MinhagenRMP}

Our numerical algorithm is based on the assumption that the time evolution of the vortex matter can be broken into two distinct parts: motion of the vortices and antivortices, and  creation of v-av pairs. The first is governed by the Langevin equations
\begin{equation}
\eta \frac{d{\bf r}_j(t)}{dt} = -{\bf{\nabla}}_j E + {\bf F}_j + {\bf{\Gamma}}_j \;,
\label{eq.lgv}
\end{equation}
where $\eta$ is the Bardeen-Stephen friction coefficient, $j$ runs over all vortices and anti-vortices present at time $t$,  ${\bf F}_j=(q_j\phi_0/c){\bf J}\times \hat{\bf z}$ is  the driving force, and ${\mbox{\boldmath $\Gamma$}_j}$ is the random force appropriate to temperature $T$. ${\bf J}$ is the applied current density which makes and angle $\alpha$ with the positive $y$-axis, as shown in Fig.~\ref{fig.system}(b). These equations are solved numerically by a finite difference method. Eventually, a vortex and an antivortex collide and annihilate each other. This is taken into account by assuming that such a collision occurs every time the vortex and the antivortex are separated by a distance $\xi_c$. 

The procedure for the second part, v-av pair creation, is to freeze the vortex matter configuration after every Langevin dynamics time step and create v-av pairs using a Monte Carlo  procedure. This assumes that the actual nucleation process takes place in a time interval so short that the motion of the vortex matter during it can be neglected. Trial v-av pairs are then placed at every point of an auxiliary grid, with unit cell of dimensions $a_c\times a_c$ ($a_c\simeq\xi_c$), commensurate with the dipole array. The vortex and antivortex are placed at a distance $\xi_c$ from each other at a given axis, with the pair center of mass fixed at the grid point. At high $T$, this axis is chosen randomly, whereas at low $T$ it is determined by the direction of the total current density calculated at the grid point (at that instant of time) as to favor pair unbinding. Then, a Metropolis code is used to either accept or reject the pair. The energy entering this algorithm includes the energy due to the interaction with the transport current, that is  $E_T = E - \sum_j {\bf F_j}\cdot{\bf r}_j$. Such a procedure accounts for both deterministic (induced by the local current distribution) and thermal pair nucleation.~\cite{Unpub}

\begin{figure}[b,t]
\centerline{\includegraphics[scale=0.90]{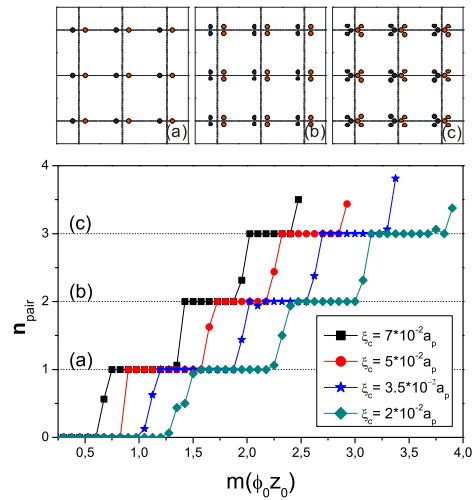}}
\caption{(Color Online) (Top) Vortex(black)-antivortex(red) configurations for (a) $m=1.0 \phi_{0} z_{0}$ ($n_{pair}=1$), (b) $m=1.65 \phi_{0} z_{0}$ ($n_{pair}=2$) and (c) $m=2.25 \phi_{0} z_{0}$ ($n_{pair}=3$). (Bottom) Number of v-av pairs per dipole, $n_{pair}$, versus magnetic moment of the dipoles, $m$, for different values of $\xi_{c}$.} \label{fig.2}
\end{figure}

The simulations are carried out on a $L\times L$ section of the bilayer, with $L=8a_p$, and assuming periodic boundary conditions in both $x$ and $y$ directions. The time step used in the numerical integration is $dt=2\cdot 10^{-4}t_0$ ($t_0=\phi_0\epsilon_0/(\eta{a_p})$) and we chose $z_0=0.2a_p$. Equilibrium configurations of the vortex matter for small $T$ and ${\bf J}=0$, were obtained minimizing $E$ (Eq.~\ref{eq.et1}) by means of a simulated annealing scheme. Fig.~\ref{fig.2}
shows the equilibrium states of the superconducting film obtained by this method. The bottom panel presents the number of vortex-antivortex pairs per dipole ($n_{pair}$) as a function of the magnetic moment of the dipoles for several values of $\xi_c$. These curves are characterized by plateaus at integer $n_{pair}$. Examples of stable vortex-antivortex configurations for $\xi_c=7\times 10^{-2} a_p$ are depicted in the upper panels. Notice that by increasing $\xi_c$ (for instance, by increasing the temperature) the magnetic moment necessary for the creation of a new v-av pair per dipole decreases. This is consistent with the fact that the force necessary to separate a v-av pair decreases with $\xi_c$. Therefore, for lower $\xi_c$ values, a strong magnetic moment is necessary to hold the pair from annihilating each other. A similar temperature dependence of the critical magnetic moment has been also observed in numerical calculation for magnetic dot arrays in the off-plane geometry\cite{Milosevic0405}.

\begin{figure}[b,t]
\centerline{\includegraphics[scale=0.80]{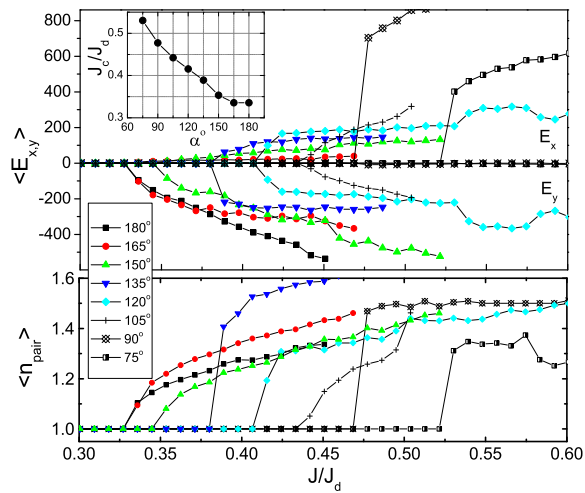}}
\caption{(Color online) Time averages of the $x$ and $y$ components of the electric field (top) and of the v-av pair density (Bottom) as functions of the current density applied at different orientations $\alpha$. Inset: angular dependence of the critical current, $J_c(\alpha)$.}
\label{fig.VI}
\end{figure}

Now we discuss the response of the vortex matter to an applied current flow. For simplicity, we shall consider hereafter only the cases where each dipole stabilize one v-av pair in equilibrium, by assuming $\xi_c=7 \times 10^{-2} a_p$ and $m=1.0\phi_0 z_0$. Fig.~\ref{fig.VI} (a)
presents the $x$ and $y$ components of the mean electric field induced by vortex motion, expressed by the time average of ${\bf E} = \phi_0 \sum{q_i\dot{{\bf r}_i}}\times\hat{z}$, as a function of the current density {\bf J} (in units of the depairing current $J_d$) for different orientations of {\bf J}. Averages were taken over $15\cdot 10^6$ time steps after a relaxation waiting time of $5\cdot 10^6$ time steps. The $E_x(J)$ and $E_y(J)$ curves evidence a strong dependence of the v-av dynamics on the current orientation $\alpha$. Such anisotropy becomes clearer when one plots the critical current $J_c$ as a function of $\alpha$. Vortices are depinned more easily when {\bf J} is at an angle $\alpha=180^\circ$. At this direction, ${\bf J}=J\hat{y}$ induces a Lorentz force on the vortices (antivortices) parallel to $+\hat{x}$ ($-\hat{x}$) in such a way as to separate the v-av pairs. Such separation is favored by the supercurrent generated right below each dipole, which is also parallel to $-\hat{y}$ and is responsible for the stabilization of the v-av pairs. At $\alpha = 90^\circ$, on the other hand, the applied current has to overcome alone the v-av mutual attraction, with no help from the dipole-induced supercurrent. For $\alpha<90^\circ$, the component $J_y$ inverts sign and tends to bind the v-av pair. At $\alpha=75^\circ$, $J_x$ is still strong enough to unbind the pairs. At lower $\alpha$, however, all v-av pairs annihilate each other and the current necessary to create new pairs and stablish a dynamical state of moving v-av matter is close to $J_d$. At such current range, our model is no longer valid and therefore $J_c$ could not be estimated for $\alpha<75^\circ$. Fig.~\ref{fig.VI} (b) presents the mean v-av pair density $\langle n_{pair}\rangle$ as a function of {\bf J}. Interestingly, for all angles studied the onset of v-av matter motion is accompanied by an increase in $\langle n_{pair}\rangle$ with respect to the static (equilibrium) value $\langle n_{pair}\rangle=1$. 

\begin{figure}[b,t]
\centerline{\includegraphics[scale=0.83]{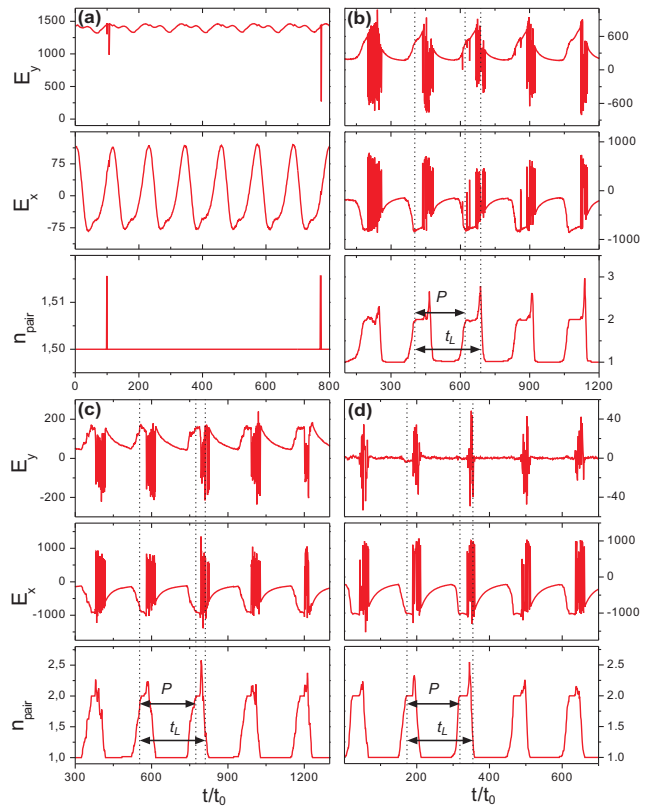}}
\caption{Time series of the electric field $E_x$, $E_y$ and v-av pair density $n_{pair}$ for $\alpha=90^\circ$ (a), $120^\circ$ (b), $150^\circ$ (c) and $180^\circ$ (d).} \label{fig.output}
\end{figure}

\begin{figure}[b,t]
\centerline{\includegraphics[scale=0.65]{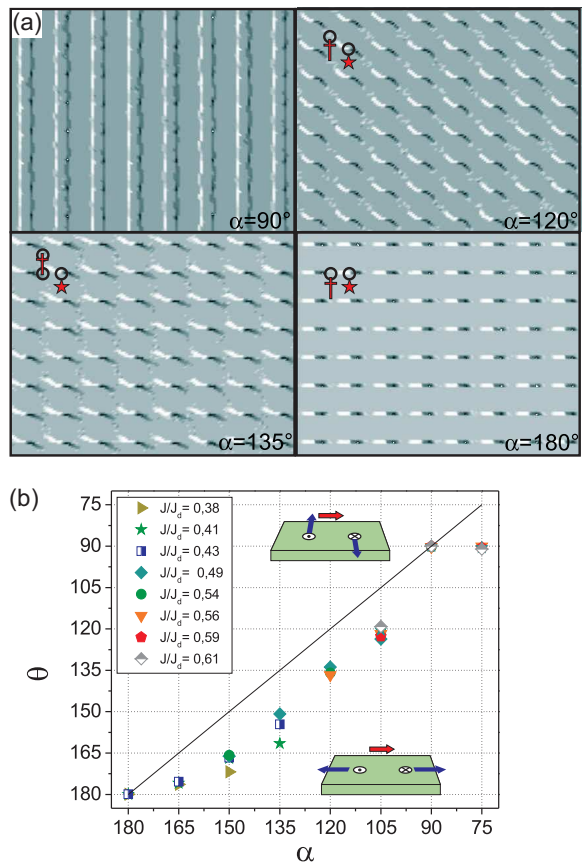}}
\caption{(Color online) (a) Time-averaged vortex density for $J=0.5J_d$ and different current orientations $\alpha$. Light (dark) shades correspond to vortices (antivortices). The circles indicate creation ($\star$) and annihilation ($\dag$) of v-av pairs. (b) Electric field direction $\theta$ as a function of $\alpha$ for several driving current intensities (symbols). The  $\theta=\alpha$ line is shown for comparison. The cartoons illustrate the dominating v-av motion for $\theta\sim 180^\circ$ and $\theta\sim 90^\circ$.} \label{fig.guide}
\end{figure}

Here we analyze in more detail the dynamics of vortices and antivortices in the film and how it reflects on the macroscopic electrical response and the mean pair density. We shall restrict ourselves to four directions, $\alpha=180^\circ$, $\alpha=150^\circ$, $\alpha=120^\circ$ and $\alpha=90^\circ$. Fig.~\ref{fig.output}
present the time evolution of the number of v-av pairs per dipole and the $x$ and $y$ components of the mean electric field for fixed ${\bf J}$ values. For $J>J_c(\alpha)$, we observed steady states of the moving v-av matter characterized by an oscillatory time dependence of $E_x$ and $E_y$, even though the applied current is constant. For $\alpha = 90^\circ$, $n_{pair}(t)$ rapidly reaches 1.5 and keeps essentially constant at this value. The corresponding steady state is illustrated in Fig.~\ref{fig.guide}(a), 
where we present contour plots of the time-averaged vortex-antivortex density, defined as $\rho({\bf r}) =  \langle\sum_iq_i \delta({\bf r-r}_i(t))\rangle_t$. The average is taken at a fixed ${\bf J}$ over $N_S$ time steps. The trajectories of vortices (antivortices) are given by the light (dark) color shades. As it is clear, vortices and antivortices follow distinct tracks, perpendicular to the applied current, and collide very rarely.

For $\alpha > 90^\circ$, the dynamical steady states are much reacher. $n_{pair}(t)$ is also oscillatory, indicating a cyclic series of creation and annihilation of v-av pairs. The general picture is as follows: when a current $J>J_c(\alpha)$ is applied, the vortices and antivortices stabilized by the dipoles in the equilibrium ($J=0$) state, which we shall refer to as first-generation vortices, depin and move apart from each other, leaving space for the creation of new (second-generation) v-av pairs right below the dipoles. At this point, $n_{pair}(t)$ reaches twice its original value ($n_{pair}(t)=2$). These new pairs, give an additional push to the preexistent ones, inducing a sudden increase in $|{\bf E}|$. All first-generation vortices and antivortices run towards annihilation, which will take place at the midpoint between first-neighbor dipoles, for $135^\circ\leq\alpha \leq 180^\circ$, or at the face center of the dipole array unit cells, for $105^\circ\leq\alpha <135^\circ$, as revealed by the $\rho({\bf r})$ plots presented in Fig.~\ref{fig.guide}(a). Just before annihilation, however, new v-av pairs spawn at the above mentioned positions and subsequently annihilate the first-generation vortices and antivortices. We believe these extra pairs arise from the fact that our model assumes a rigid vortex core. At distances of order $\xi$, the v-av annihilation in superconductors is characterized by strong vortex core deformation. In our model, the extra v-av pairs appear naturally as a way of compensating the vortex core rigidity and accelerate the annihilation process. As a side effect, a spurious, high-frequency noise arises in $E_x(t)$ and $E_y(t)$. Notwithstanding, we have checked that such noise has only a neglectable influence on the macroscopic quantities shown in Fig.~\ref{fig.VI}. Finally, after the annihilation process, the second-generation v-av pairs become first-generation ones and the whole process repeats. Notice that the lifetime, $t_{\rm L}$, for a vortex of the first or second generation is always larger than the oscillation period $P$ (see Fig.~\ref{fig.output}).

The angular dependence of the v-av dynamics described above  suggests that motion occurs preferentially along the $x$ directions for $\alpha$ close to 180$^\circ$. Such guided motion of the v-av matter should reflect directly on the electric field {\bf E} generated by vortex motion, giving rise to a deflection of {\bf E} with respect to {\bf J} and thus inducing transverse resistance. To demonstrate this statement, we plot the direction of ${\bf E}$, $\theta$, as a function of the direction of ${\bf J}$, $\alpha$, for several values of $J$ (left panel of Fig.~\ref{fig.guide}). Indeed, $\theta(\alpha)$ diverges considerably from the $\theta=\alpha$ line. For $\alpha$ close to $180^\circ$, $\theta$ tends to lock at $\theta \simeq 180^\circ$ in the whole $J$ range, whereas for $\alpha=75^\circ$, {\bf E} locks at $\theta=90^\circ$. This transverse resistance resulting from the guided motion of v-av matter is essentially different from that observed in superconducting films with non-magnetic pinning arrays\cite{guidance} and other Hall-like effects in the sense that here no macroscopic magnetic field is required.

In summary we have studied the dynamics of vortices and antivortices in superconducting films interacting with an array of in-plane magnetic dipoles by means of a model which explicitly takes into account creation and annihilation of v-av pairs. The current-driven dynamics of the vortex-antivortex matter at zero applied field is characterized by a series of creation and annihilation processes producing an oscillatory behavior of the time dependent electric field. Other predictions of our model include the anisotropy in the critical current and the transverse electric field induced by vortex-antivortex guidance. These results can be checked out experimentally by means of conventional transport measurements at zero applied magnetic field or by direct high-resolution magneto-optical imaging. We anticipate that the zero-field guidance effect should also occur for other dipole polarizations, such as the off-plane geometry, as long as the magnetic texture captures v-av pairs. 

\acknowledgments The authors are grateful to Gilson Carneiro for fruitful discussions. Research supported in part by the Brazilian agencies CNPq and CAPES.

\end{document}